\documentclass[runningheads]{llncs}
\usepackage{times}
\usepackage{epsfig}
\usepackage{graphicx}
\usepackage{amssymb}
\usepackage{pifont}
\usepackage{amsmath}
\usepackage{hyperref}
\usepackage{adjustbox}
\usepackage{caption}
\usepackage{booktabs}
\usepackage{amsmath}
\usepackage{amssymb}
\usepackage{dsfont}
\usepackage{multirow}
\usepackage{lipsum}
\newcommand*\samethanks[1][\value{footnote}]{\footnotemark[#1]}
\def\blfootnote{\gdef\@thefnmark{}\@footnotetext}
\makeatother

\begin{document}
\title{Multi-Contrast Computed Tomography Atlas of Healthy Pancreas}
\author{
Yinchi Zhou\inst{1}\thanks{Contributed equally to this work.} \and
Ho Hin Lee\inst{1}\samethanks \and
Yucheng Tang\inst{5} \and
Xin Yu\inst{1} \and
Qi Yang\inst{1} \and
Shunxing Bao\inst{1} \and
Jeffrey M. Spraggins \inst{3} \and
Yuankai Huo\inst{1,2} \and
Bennett A. Landman\inst{1,2,4}}
%

\institute{Anonymous}

\institute{Department of Computer Science, Vanderbilt University \and
Department of Electrical and Computer Engineering, Vanderbilt University \and
Department of Biochemistry, Vanderbilt University \and
Department of Radiology, Vanderbilt University Medical Center \and
NVIDIA }
\maketitle              
\begin{abstract}
With the substantial diversity in population demographics, such as differences in age and body composition, the volumetric morphology of pancreas varies greatly, resulting in distinctive variations in shape and appearance. Such variations increase the difficulty at generalizing population-wide pancreas features. A volumetric spatial reference is needed to adapt the morphological variability for organ-specific analysis. Here, we proposed a high-resolution computed tomography (CT) atlas framework specifically optimized for the pancreas organ across multi-contrast CT. We introduce a deep learning-based pre-processing technique to extract the abdominal region of interests (ROIs) and leverage a hierarchical registration pipeline to align the pancreas anatomy across populations. Briefly, DEEDs affine and non-rigid registration are performed to transfer patient abdominal volumes to a fixed high-resolution atlas template. To generate and evaluate the pancreas atlas template, multi-contrast modality CT scans of 443 subjects (without reported history of pancreatic disease, age: 15-50 years old) are processed. Comparing with different registration state-of-the-art tools, the combination of DEEDs affine and non-rigid registration achieves the best performance for the pancreas label transfer across all contrast phases (non-contrast: 0.497, arterial: 0.505, portal venous: 0.494, delayed: 0.497). We further perform external evaluation with another research cohort of 100 de-identified portal venous scans with 13 organs labeled, having the best label transfer performance of 0.504 Dice score in unsupervised setting. The qualitative representation (e.g., average mapping) of each phase creates a clear boundary of pancreas and its distinctive contrast appearance. The deformation surface renderings across scales (e.g., small to large volume) further illustrate the generalizability of the proposed atlas template.

\end{abstract}
%
%
%




\section{Introduction}
\label{Introduction}
With the complicated relationship between physiological and metabolic process in the human body, substantial efforts are underway to map the organization and molecular profiles of cells within specific tissues \cite{rozenblatt2017human}. Adapting multi-scale context from cell to organ level is also essential to provide a better definition to correlate the biomarkers across different imaging domains \cite{hubmap2019human}. Computed tomography (CT) is widely used to visualize the patients' anatomy at a system scale \cite{brenner2007computed}. While CT provides the anatomical correspondence in a system-scale only, a standardized framework is needed to generalize information across scales and enhances the ability of visualizing the complex organizations of tissues across populations.

\begin{figure*}[t!]
    \centering
    \includegraphics[width=\textwidth]{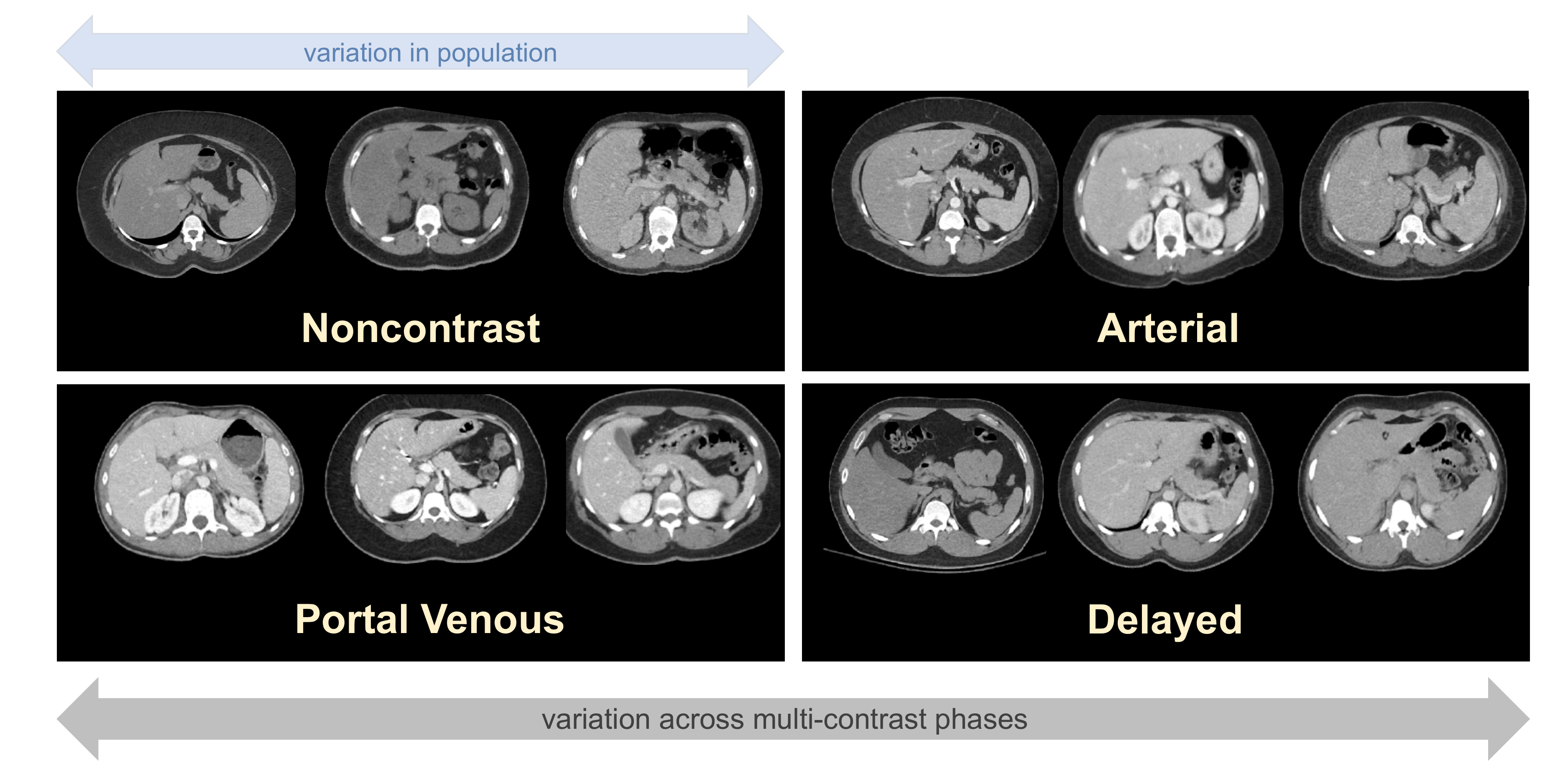}
    \caption{The anatomical characteristics of pancreas organs vary widely in population. Visual difference between each contrast phase is also shown a) noncontrast b) arterial c) portal venous d) delayed. A generalizable reference is needed to adapt individual morphological variability within contrast phases. }
\end{figure*}

During the imaging procedures, contrast enhancement is performed by injecting a contrast agent before imaging. Multi-phase contrast enhanced CT are generated with the sharpened anatomical and structural details between neighboring organs for diagnosis \cite{rusko2009automatic,nayak2019computer,tang2021phase}. Four different phases are typically generated according the retaining time of contrast agent in the imaging cycle: 1) non-contrast, 2) arterial, 3) portal venous, and 4) delayed. With the variation in contrast level between organs, the large range of intensity levels are beneficial to capture the fine-grained contextual features of each specific organ, especially for the pancreas organ. By visualizing the anatomical context of pancreas across large population cohorts, we observe that the volumetric morphology of healthy pancreas organ varies with respect to demographics (e.g., sex, body size), as shown in Fig. 1. To investigate the population-wise healthy biomarkers of pancreas organ in the systemic level, a standardized imaging atlas framework is needed to adapt the population-wise anatomical characteristic onto one single template with image registration technique. However, with the large variation in organ anatomies and body sizes across population, generating such standard reference template for pancreas organ is still challenging and no atlas framework for pancreas organ is currently available in public.

Creating an organ/tissue-specific atlas framework has been widely leveraged with magnetic resonance imaging (MRI) and extensive efforts have been applied to leverage brain MRI to investigate biomarkers for multiple perspectives \cite{wang2020allen,lorenzen2006multi}. Due to the similarity between human brain and mouse brain, Kova{\v{c}}evi{\'c} et. al proposed a 3D variational atlas with mouse brain to represent the average anatomy and the variation among the population \cite{kovavcevic2005three}. \textit{Wang} et.al created a population average reference framework leveraging 1675 specimens of mouse brain MRI \cite{wang2020allen}. On the other hand, \textit{Shi} et al. created an unbiased infant brain atlas with group-wise registration from three different scanning time points using MRI from 56 males and 39 females \cite{shi2011infant}. \textit{Kuklisova-Murgasova} et al. proposed multiple atlas to generalize the aging characteristics from 29 to 44 weeks infants \cite{kuklisova2011dynamic}. \textit{Ali} et al. generates an unbiased spatial-temporal 4-D atlas and time-variable longitudinal atlas for infant brain \cite{gholipour2017normative}. To further investigate into the aging characteristics in brain tissue, \textit{Yuyao} et al. leveraged patch-based registration in spatial-temporal wavelet domain to generate longitudinal atlas \cite{zhang2016consistent}. While previous efforts was concentrated on generating healthy brain atlas template, \textit{Rajashekar} et al. proposed high-resolution normative atlases to visualize the population-wise representation of brain disease (e.g., brain lesion, stroke) in both FLAIR MRI and non-contrast CT modalities. For abdominal regions, pioneer studies have been demonstrated to develop a multi-contrast kidney atlas that generalize both contrast and morphological characteristic within kidney organs \cite{lee2021construction,lee2022multi}. Furthermore, such kidney atlas template is further extended to generalize the substructure organ (e.g., medulla, renal context, pelvicalyceal systems) in kidney regions with arterial phase CT \cite{lee2022supervised}. However, limited studies have proposed to create a standard reference atlas for pancreas organs with its challenging morphology associated with patients' demographics.

In addition to putting efforts into generating tissue/organ atlases, robust image registration algorithms to transfer the anatomical context onto one single template are also vitally important. Previous works have sought to enhance the registration performance by innovating conventional frameworks with affine and deformable transformation \cite{ashburner2007fast,avants2008symmetric,balakrishnan2018unsupervised}. The spatial transformation is optimized by regularizing the deformation field to align the anatomical context from moving image to a single fixed template with traditional approach such as discrete optimization \cite{dalca2016patch}, b-spline deformation \cite{rueckert1999nonrigid}, Demons \cite{vercauteren2009diffeomorphic}, and symmetric normalization \cite{avants2008symmetric}. To further enhance the efficiency and robustness of registration algorithm, deep learning was introduced to generate large deformation field by extracting meaningful representations for deformation field predictions. VoxelMorph is a pioneering network that optimizes with a generalized function to compute deformation field in unsupervised setting \cite{balakrishnan2019voxelmorph,dalca2019unsupervised}. While VoxelMorph was initially optimized with brain imaging, a large deformation field is needed for abdominal imaging due to the significant variation in body size and organ morphology across demographics. \textit{Zhao} et al. adopt Voxelmorph framework and extended as a recursive cascaded network that leverage organ labels to crop the organ-specific region of interests (ROIs) and progressively registered the anatomical context to the fixed template \cite{zhao2019recursive}. \textit{Zhao} et al. introduced a deep learning framework that generates bounding boxes to initially localize multiple organ ROIs and leverage the organ-specific patches for registration \cite{zhao2021target}. However, previous approaches only demonstrate the feasibility of registering organ-specific ROIs. \textit{Heinrich} et al. adopt substantial deformation in compute volumetric scans by innovating a probabilistic dense displacement network with organ label supervision \cite{heinrich2019closing}. Yet, voxel-wise labels are needed to supervise the training process and instability in performance may exist due to the domain shift with unseen data \cite{lee2022multi}.

\begin{figure*}[t!]
    \centering
    \includegraphics[width=\textwidth]{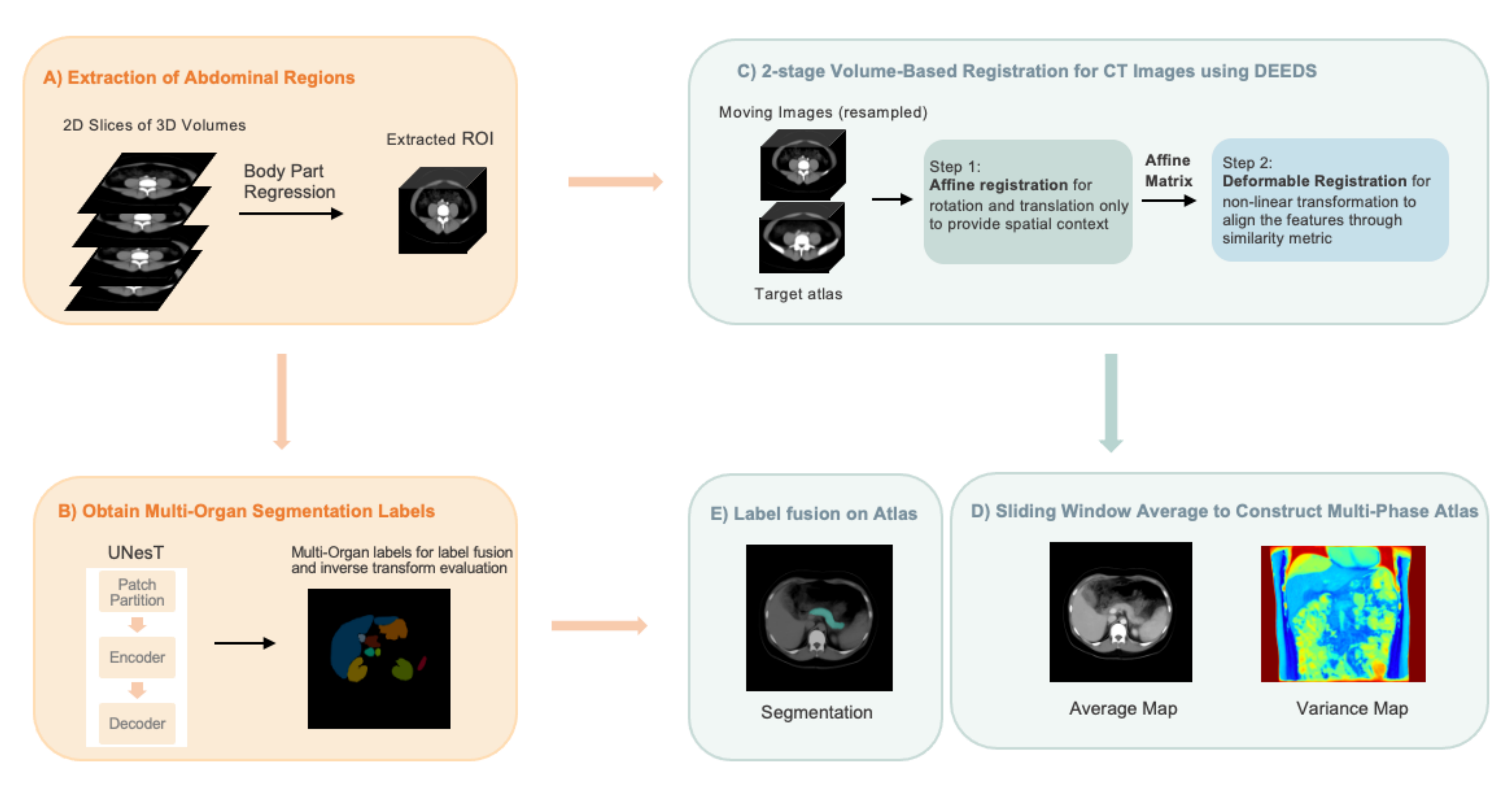}
    \caption{Complete overview of our propose atlas generation framework. We extract the abdominal regions from CT scans using the body part regression network. We register the cropped ROI to the reference image with a hierarchical two-stage registration and compute both average and variance mappings to evaluate the effectiveness of the atlas framework. Furthermore, we statistically fuse the pseudo predictions from the transformer-based segmentation network UNesT and perform inverse transformation back to the subject space for evaluation.}
\end{figure*}

In this work, we propose a high-resolution CT pancreas atlas framework that optimized for the healthy pancreas. Briefly, we initially crop the abdominal ROI from each each subject scan with a deep neural network called body part regression (BPR), which aims to minimize the field of view (FOV) differences between scans and reduces the failure rate of registration. Specifically, we slice the volumetric scans and input each slice into BPR network to compute a regressed score ranging from -12 to +12, referring as the upper lung region to the pelvis region in the body. We then limit the predicted score range and extract the ROI within abdominal region only. After data preprocessing, a two-stage hierarchical registration pipeline is performed to register each subject scan to the high-resolution atlas target with metric-based registration across all contrast phases \cite{heinrich2013towards,heinrich2015multi}. To evaluate the quality of anatomical transfer across scans, we compute average and variance mappings to demonstrate the organ appearance across all registered outputs in each phase and further quantify the registration performance with inverse label transfer from atlas framework. Overall, our main contributions are summarized as four folds:


\begin{itemize}
\item[\textbullet] We established a standardized framework to obtain a population-based pancreas atlas.

\item[\textbullet] We propose a hierarchical metric-based framework optimizing for healthy pancreas to generalize the anatomical and contrast characteristics of pancreas organ across demographics and domain shift of contrast phases.

\item[\textbullet] We evaluate the effectiveness of our proposed atlas template by inversely transforming the atlas target labels to a de-identified research cohort of 100 abdominal scans with 13 organs ground-truth labeled. A large population of unlabeled multi-contrast phase CT cohort is leveraged to compute average and variance mappings to demonstrate the generalizability of the atlas framework. Our proposed atlas framework achieves a stable transfer ability in pancreas organ with an average Dice of 0.504 in unsupervised setting.

\item[\textbullet] The average template generated, and the associated pancreas organ labels is available public for usage through HuBMAP.
\end{itemize}

\section{Methods}
\subsection{Self-Supervised Body Part Regression Network for Preprocessing}
With the substantial variation in imaging protocols, the imaging samples from a large cohort usually present with different range of field of views (FOVs). Such variability of FOV may increase the failure rate of registration when the FOV difference between the subject scan and the atlas template is large. Here, we adapt body part regression (BPR) network to extract similar FOV within the abdominal regions only, thus to enhance the registration performance. Specifically, given an unlabeled dataset $\{{x_u}_i\}_{i=1}^N$ as the moving image domain, and the atlas image $\{x_a\}$, our goal is to crop the volumetric $x_u$ to have an approxiamte FOV of abdominal interest only with the atlas template $x_a$. \textit{Tang} et al. proposed a self-supervised BPR network to generate a continuous score for each axial slice in the volumetric scans as the normalized body localization values \cite{yan2018unsupervised,tang2021body}. Each score is within the range of -12 to + 12, which refers to different anatomical location (e.g., -12: upper chest, -5: diaphragm/upper liver, 4: lower retroperitoneum, 6: pelvis). For abdominal region, we limit the score ranges across all axial slices within -5 to 5 and crop the slices.

\subsection{Transformer-Based Segmentation Network}
Apart from data preprocessing, pancreas organ labels are needed to further quantify the healthy biomarkers across populations and perform statistically label fusion to generate the atlas label. With the recent advance of deep neural networks in medical image segmentation \cite{tang2021high,lee2021rap,lee20223d}, transformer-based networks have been further proposed to leverage with respect to their proficiency in capturing long-range dependency and the large receptive filed behavior \cite{hatamizadeh2021unetr,hatamizadeh2022swin}. Here, we adapt a transformer-based model that incorporates a hierarchical design alongside a block aggregation method, attaining state-of-the-art performance across multiple modalities and public datasets \cite{yu2022unest}. This model initially projects 3D volumes into a sequence of patches, and then employs a 3D block aggregation algorithm to augment communication between these patches. Within each hierarchical level, patches are transformed into blocks and introduced to the transformer layer. The output then undergoes the block aggregation algorithm and is subsequently downsampled for the next hierarchy. The model consists of three hierarchies with a configuration of 64, 32, and 1 block(s) respectively. The block aggregation algorithm leverages local attention, therefore enhancing data efficiency. In our study, the needs for segmentation labels on abdominal scans is twofold: for inverse transform evaluation and joint label fusion to visualize the pancreas region on atlases.


\begin{figure}[t!]
    \centering
    \includegraphics[width=\textwidth]{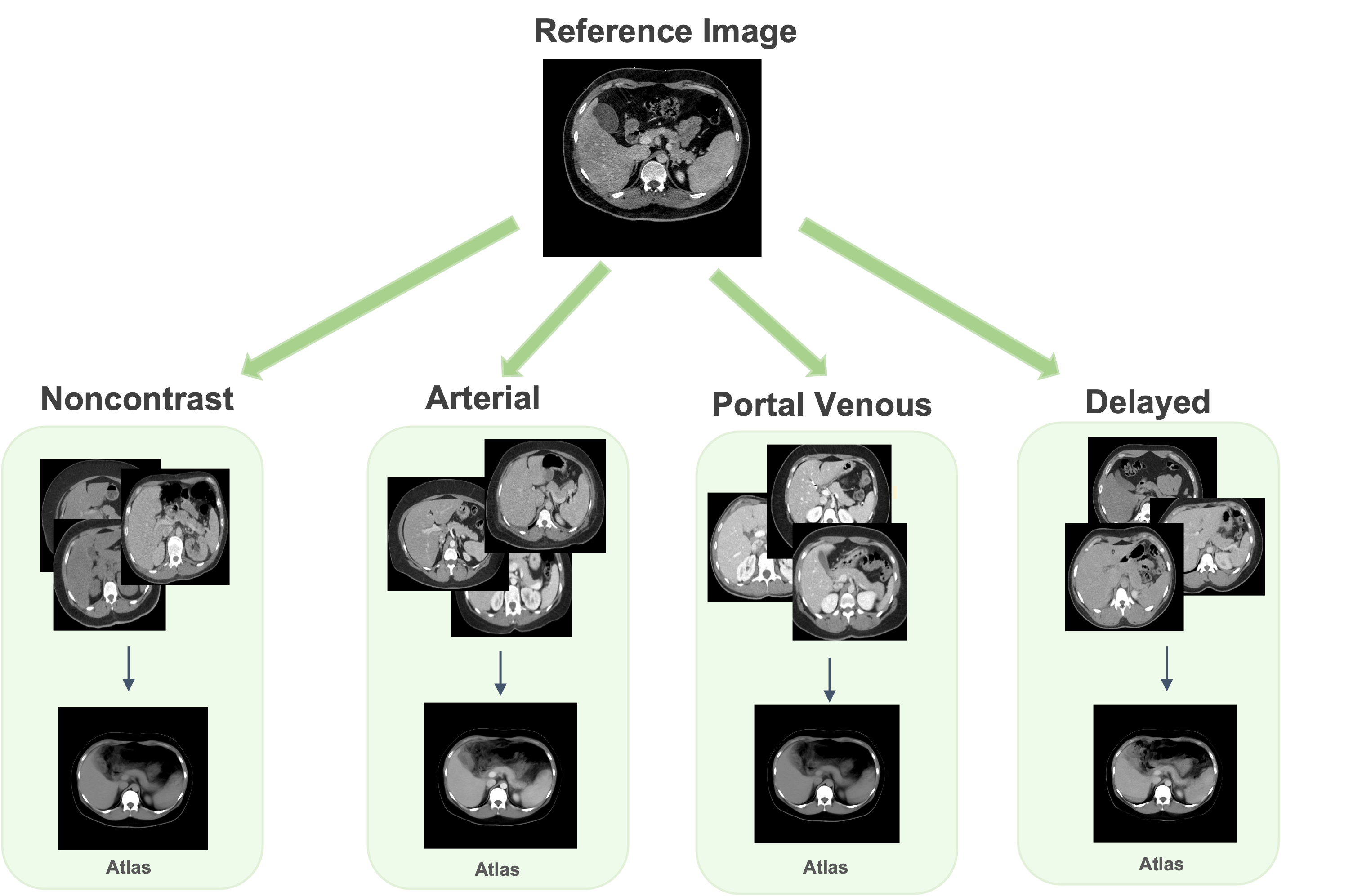}
    \caption{To evaluate the effectiveness of our atlas framework, inverse transformation is performed to backpropagate the anatomical label from the atlas template to the moving subject domain and quantify the similarity with the corresponding ground truth label.}
\end{figure}

\subsection{2-Stage Hierarchical Registration}
Our registration framework consists of two main steps: 1) affine registration and 2) deformable registration. Inspired by \textit{Lee} et al. \cite{lee2022multi,lee2022supervised}, we adapt dense displacement sampling (DEEDS) as our registration backbone for abdominal imaging. The concepts of DEEDS is to compute a large deformation filed with a discretized sampling space to align the anatomical context across andominal organs with variable morphology \cite{heinrich2013towards,heinrich2015multi,heinrich2013mrf}. We first perform DEEDS affine registration to align the abdominal organs with 12$\circ$ degrees of freedom from moving images to the atlas template, coarsely providing prior spatial information and each affine component with the generated affine transformation matrix as our first stage output. The affine-aligned intermediates are leveraged for the DEEDS deformable registration as our second stage registration process. The DEEDS deformable registration refines the spatial relationship between randomly selected patches and compute the local voxel-wise correspondence with its specific similarity metric as follows:
\begin{equation}
    D(x_m, p_x, p_y) = exp(\frac{S(p_x,p_y)}{q^2}, \:\:\:p_x,p_y\in N
\end{equation}
where $p_x$ and $p_y$ are denoted as the center coordinate of another patch from one of the neighbourhood $N$, $S$ is the self-similarity metric, which is optimized by a distance function $D$ between the image patches from the moving samples. $q^2$ denotes as a function to evaluate the noise in the local and global perspectives. The similarity metric enhance the avoidance of the adverse effect from image artifacts or random noise from the central extracted patch. During the deformable registration, five different levels are leveraged in grid spaceing ranging from eight to four voxels to randomly select patches. The displacement search radii are defined from six to two steps between five and one voxels. Six neighborhoods are chosen to compute 12 distances between pairwise patches for optimization \cite{heinrich2013towards,heinrich2015multi,heinrich2013mrf}. Both deformed scans with the corresponding displacement matrix are generated as the final output and align the fine-grain anatomical details of each organ.

\begin{figure*}[t!]
    \centering
    \includegraphics[width=\textwidth]{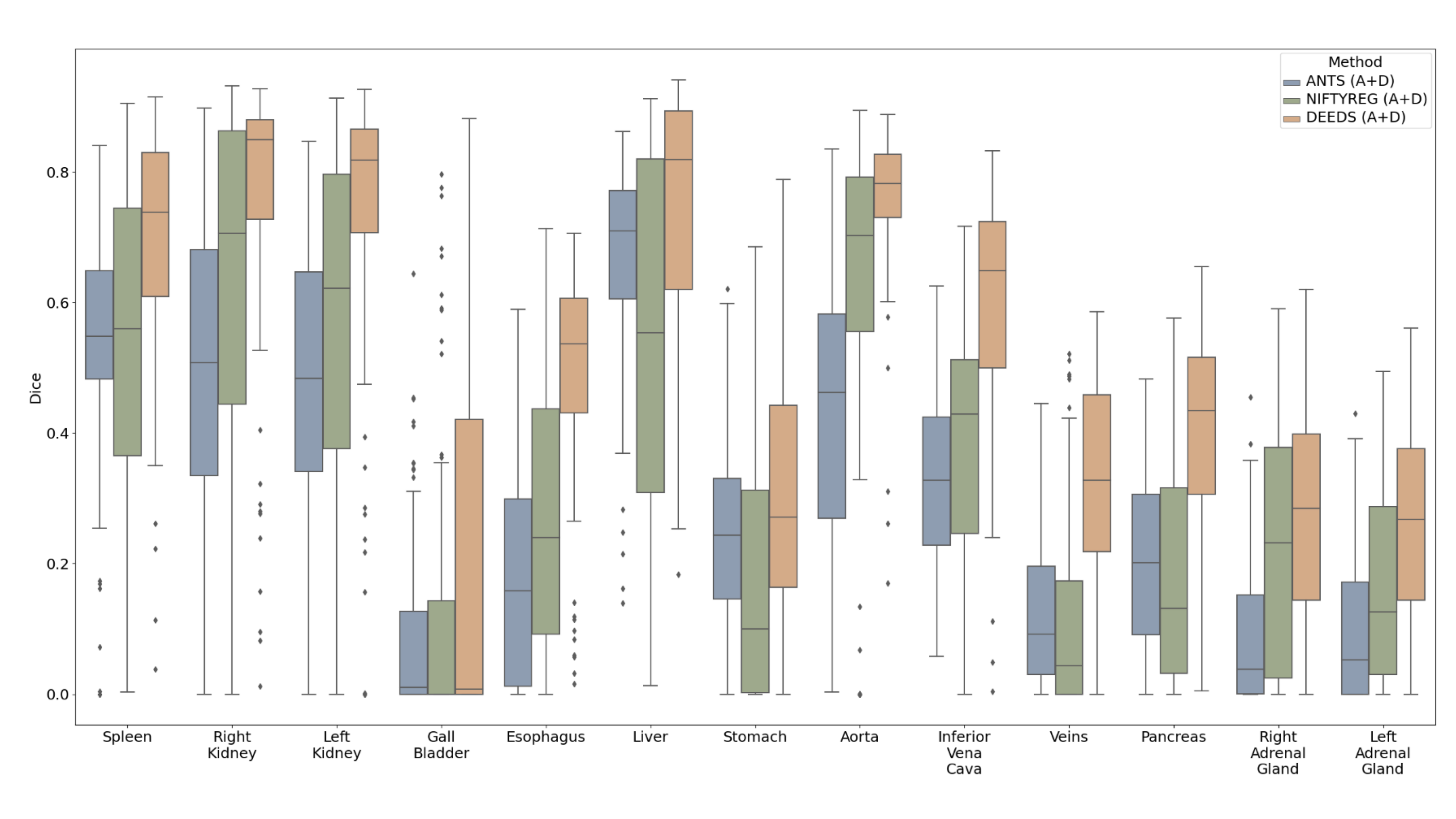}
    \caption{The quantitative representation of inverse label transfer with multi-organ portal venous CT dataset are demonstrated that DEEDS with affine registration outperforms the other two traditional methods. }
\end{figure*}
\section{Experimental Setup}
\subsection{Datasets}
\label{dataset}
\noindent\textbf{Clinical Research Multi-Phase CT Cohort:} A total of 443 multi-contrast phase CT volumes are selected for the formation of multi-phase atlases from a large cohort of abdominal CT scans of 2000 patients under the approval of Institutional Review Board (IRB \#131461). The scans include retro-peritoneal and abdominal organs. Since our goal is to construct a healthy CT atlas optimized for pancreas region, we exclude patients who exhibit pancreatic lesions. 898 subjects whose age ranging from 18 to 50 years old. After quality assessment of registered subjects, 443 unlabeled CT volumes were used to create the atlas for four phases: 59 volumes for non-contrast phase, 40 volumes for arterial phase, 330 for portal venous phase, 14 for delayed phase. All volumetric scans are initially reoriented to a standard orientation (RAS) before data preprocessing \cite{jenkinson2012fsl}. BPR is then performed to each scan and obtain the FOV in abdominal region only. Each scan is then resampled to the same resolution and dimension of the atlas template for registration. \par
\noindent\textbf{Multi-Organ Labeled Portal Venous Abdominal CT Cohort:} To evaluate the generalizability of our proposed atlas framework, we further leverage a separate healthy clinical cohort with 100 de-identified portal venous phase abdominal CT scans and 20 of the volumetric scans are the testing scans in the MICCAI 2015 Multi-Atlas Abdomen Labeling (BTCV) challenge. All volumetric scans are labeled with 13 multiple organs including: 1) spleen, 2) right kidney, 3) left kidney, 4) gall bladder, 5)
esophagus, 6) liver, 7) stomach, 8) aorta, 9) inferior vena cava (IVC), 10) portal splenic vein (PSV), 11) pancreas, 12) right adrenal gland (RAD), 13) left adrenal gland (LAD). \par
\noindent\textbf{High-Resolution Single Subject Atlas Template:} Inspired by \cite{lee2022multi}, we choose the atlas template under several conditions: 1) high-resolution in both in-plane and through-plane, 2) distinctive contrast appearance in pancreas organ morphology with clear boundary, and 3) in healthy condition. The atlas template is provided by Human Biomolecular Atlas Program (HuBMAP) with a high resolution of $0.8 \times 0.8 \times 0.8$ and the corresponding dimension of $512 \times 512 \times 434$. The atlas template is annotated with 13 organs by performing statistical label fusion with the pseudo segmentation from all registered subjects.

\subsection{Implementation Details}
For the BPR model, we pretrain a U-Net like architecture network with a total of 230,625 2D slices from a large population samples of 1030 whole body CT scans (collected from public domain) \cite{tang2021body}, while the multi-organ labeled portal venous phase CT are leveraged as external validation only. The pretrained U-Net model is end-to-end optimized with Adam optimizer with a learning rate of 0.0001 and batch size of 4. \par
After we preprocess all imaging samples with BPR model, we further evaluate our proposed atlas template in multiple perspectives. We first investigate the effectiveness of current state-of-the-art registration tools for abdominal imaging across all contrast phases in both qualitative and quantitative perspective. We performed extensive analysis with tools such as ANTS \cite{xu2016evaluation,avants2008symmetric}, NIFTYREG \cite{xu2016evaluation,modat2010fast}, and DEEDS \cite{heinrich2013mrf,heinrich2013towards,heinrich2015multi} as metric-based registration with multi-organ labeled portal venous CT cohort. Before the registration, all preprocessed scans are upsampled to the same resolution with the atlas target and perform the 2-stage hierarchical registration in high-resolution setting. We further perform multiple ablation studies to investigate the effectiveness of the BPR preprocessing to enhnace the successful rate of abdominal organ registration. Moreover, we perform ablation study to search the optimal range of BPR score to define the best FOV correlation between all moving images and the atlas target.

\subsection{Evaluation Metrics}
We adapt two commonly used metrics to evaluate the similarity between the inverse transferred labels from atlas space to moving subject space and the corresponding moving ground truth label: 1) Dice similarity score, and 2) Hausdorff distance (HD). The definition of Dice score is to compute the overlapping ratio between the predicted pixel/voxel-wise label and the ground truth label. The Dice score is defined as follows:
\begin{equation}
    Dice(P,G)=\frac{2|P\cap G}{|P|+|G|}
\end{equation}
where $P$ denotes as the label prediction and $G$ is the corresponding ground truth label, while $\|$ denotes as L1 normalization. For computing the Hausdorff distance, we extract the 3-dimensional coordinates of each vertice from the surface rendering of both the predicted label and the ground truth. We compute the Hausdorff distance with the vertices as follows:
\begin{equation}
        HD(v_p,v_g) = sup\:inf\:Dist(v_p,v_g)
\end{equation}
where $v_p$ and $v_g$ denotes as the vertice coordinates of the label prediction and ground-truth label respectively. $Sup$ and $inf$ refer to the upper and lower bound of the distance function $Dist$ value.

\begin{table}[t!]
\caption{Inverse label transfer performance of different registration methods on 13 organs average for Clinical Research Multi-phase CT Cohort (*: p$<$0.0001, Wilcoxon signed-rank test, A: affine registration, D: deformable registration)}
\begin{adjustbox}{width=\textwidth}
\begin{tabular}{c|cc|cc|cc|cc}
\toprule
& \multicolumn{2}{c}{Non-Contrast} & \multicolumn{2}{c}{Arterial} & \multicolumn{2}{c}{Portal Venous} & \multicolumn{2}{c}{Delayed} \\
\midrule
Methods & \:Dice\:$\uparrow$ & HD (mm)\:$\downarrow$\: & \:Dice\:$\uparrow$ & HD (mm)\:$\downarrow$\: & \:Dice\:$\uparrow$ & HD (mm)\:$\downarrow$\: & \:Dice\:$\uparrow$ & HD (mm)\:$\downarrow$\: \\
\midrule
ANTS (A) & \:0.256$\pm$0.115 & 47.9$\pm$22.1\: & \:0.242$\pm$0.103 & 39.9$\pm$16.8\: & \:0.242$\pm$0.082 & 42.2$\pm$18.0\: & \:0.249$\pm$0.078 & 42.5$\pm$20.1\: \\
NIFTYREG (A) & \:0.232$\pm$0.141 & 52.8$\pm$22.8\: & \:0.200$\pm$0.131 & 47.8$\pm$17.3\: &\:0.201$\pm$0.117 & 46.4$\pm$19.0\: & \:0.211$\pm$0.103 & 47.5$\pm$15.9\: \\
DEEDS (A) & \:0.192$\pm$0.103 & 48.7$\pm$17.8\: & \:0.144$\pm$0.112 & 46.4$\pm$12.6\: & \:0.148$\pm$0.111 & 45.3$\pm$14.5\: & \:0.150$\pm$0.094 & 49.0$\pm$16.5\: \\
ANTS (A+D) & 0.288$\pm$0.121 & 47.2$\pm$22.4 & 0.302$\pm$0.114 & 38.8$\pm$17.1 & 0.308$\pm$0.097 & 40.1$\pm$18.4 & 0.318$\pm$0.065 & 41.3$\pm$20.1 \\
NIFTYREG (A+D) & 0.314$\pm$0.154 & 51.7$\pm$22.2 & 0.334$\pm$0.180 & 45.7$\pm$16.3 & 0.350$\pm$0.176 & 45.6$\pm$19.1 & 0.378$\pm$0.150 & 46.8$\pm$17.8 \\
\textbf{DEEDS (A+D)} & \textbf{0.497$\pm$0.076*} & \textbf{39.2$\pm$21.9*} & \textbf{0.505$\pm$0.075*} & \textbf{32.5$\pm$14.7*} & \textbf{0.494$\pm$0.077*} & \textbf{33.3$\pm$17.5*} & \textbf{0.497$\pm$0.086*} & \textbf{37.1$\pm$20.7*} \\
\bottomrule
\end{tabular}
\end{adjustbox}
\end{table}

\begin{table}[t!]
\caption{Inverse label transfer performance of different registration methods on 13 organs for multi-organ labeled portal venous phase CT Cohort (*: p$<$0.0001, Wilcoxon signed-rank test, A: affine registration, D: deformable registration)}
\begin{adjustbox}{width=\textwidth}
\begin{tabular}{l|ccccccccccccc|c}
\toprule
\midrule
Methods & \multicolumn{1}{c}{Spleen} & \multicolumn{1}{c}{R. Kid} & \multicolumn{1}{c}{L. Kid} & \multicolumn{1}{c}{Gall.} & \multicolumn{1}{c}{Eso.} & \multicolumn{1}{c}{Liver} & \multicolumn{1}{c}{Stom.} & \multicolumn{1}{c}{Aorta} & \multicolumn{1}{c}{IVC} & \multicolumn{1}{c}{PSV} & \multicolumn{1}{c}{Panc.} & \multicolumn{1}{c}{RAG} & \multicolumn{1}{c}{LAG} & \multicolumn{1}{c}{Avg}\\ \midrule
\midrule
ANTS (A+D) &
0.536$\pm$0.175 & 0.498$\pm$0.226 &
0.484$\pm$0.206 &
0.091$\pm$0.139 &
0.183$\pm$0.167 &
0.662$\pm$0.158 &
0.256$\pm$0.138 &
0.431$\pm$0.201 & 
0.323$\pm$0.143 & 
0.121$\pm$0.110 &
0.204$\pm$0.132 &
0.092$\pm$0.110 & 
0.095$\pm$0.111 & 
0.306$\pm$0.246 \\
NIFTYREG (A+D)&
0.544$\pm$0.241 & 0.611$\pm$0.290 &
0.556$\pm$0.287 & 
0.113$\pm$0.206 & 
0.270$\pm$0.201 &
0.538$\pm$0.287 & 
0.175$\pm$0.198 & 
0.615$\pm$0.264 & 
0.379$\pm$0.196 & 
0.111$\pm$0.143 &
0.184$\pm$0.170 & 
0.222$\pm$0.182 & 
0.162$\pm$0.151 & 0.344$\pm$0.293 \\
DEEDS (A+D) & 
\textbf{0.697$\pm$0.178*} & 
\textbf{0.756$\pm$0.207*} & 
\textbf{0.729$\pm$0.224*} & 
\textbf{0.211$\pm$0.283*} & 
\textbf{0.491$\pm$0.166*} & 
\textbf{0.746$\pm$0.180*} & 
\textbf{0.315$\pm$0.208*} & 
\textbf{0.756$\pm$0.200.116*} & 
\textbf{0.596$\pm$0.173*} & 
\textbf{0.329$\pm$00.165*} & 
\textbf{0.398$\pm$0.168*} & 
\textbf{0.274$\pm$0.169*} & 
\textbf{0.253$\pm$0.155*} & 
\textbf{0.504$\pm$0.280*} \\
\bottomrule
\end{tabular}
\end{adjustbox}
\end{table}

\begin{figure*}[t!]
    \centering
    \includegraphics[width=\textwidth]{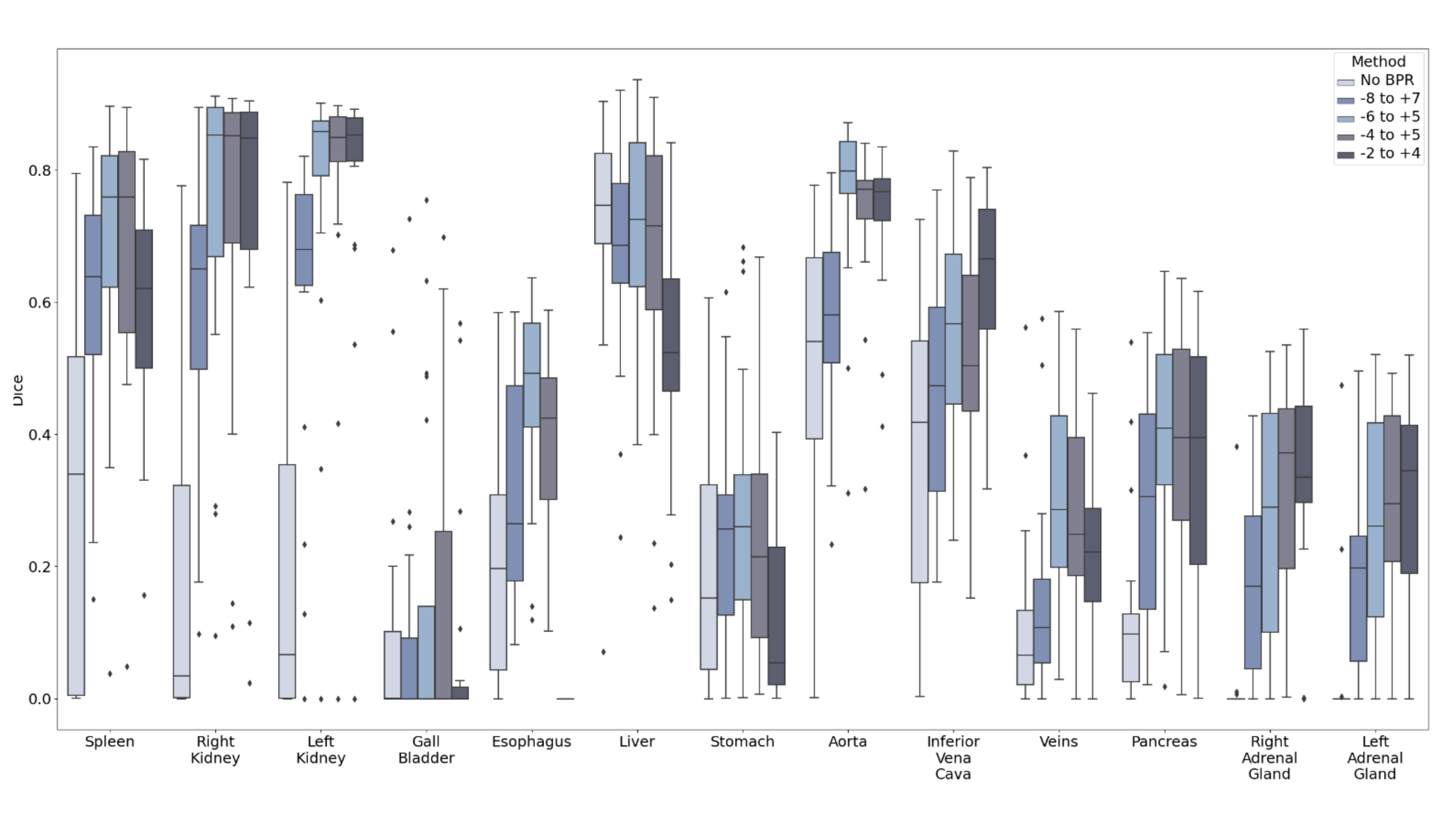}
    \caption{We perform ablation study of evaluating inverse label transfer performance with different cropping value range for body part regression. We found that the optimal range for body part regression is between -6 and +5.}
\end{figure*}
\section{Results}
\subsection{Evaluation with Clinical Research Cohort and Multi-Organ Labeled Cohort}
After cropping the abdominal regions from the raw data using body part regression, we resampled the cropped volumes and registered them to the reference volume using DEEDs. We then qualitatively assessed the registered subjects and removed those with registration failure. The average atlas for the four phases was subsequently obtained from the registered volumes. Fig. 5 presents the tri-planar view of the average atlas, in which the anatomical features of the pancreas can be clearly seen in all directions. Furthermore, the shape of surrounding organs, such as the spleen, is well-preserved, providing a better anatomical context for studying the pancreas. The variance of average templates is shown in Fig. 6. We calculate the log-scale variance between the average template and the registered subjects used to create the atlas, normalizing them to a range of 0 to 1. Among all registered regions, the pancreas exhibits relatively low variance for all phases compared to the spleen and spine. The non-contrast and portal venous phases have lower variance, approaching 0.
\par To enhance the identification and visualization of the pancreas region in the atlases, we also acquired segmentation labels by employing a label integration technique that relies on majority voting. As depicted in Figure 5, the pancreas' position and shape in the multi-phase atlases closely resemble the target reference, indicating that the anatomical and contextual information of the pancreas has been effectively transferred into the target space. Even the areas of greatest variation, such as the head and tail of the pancreas, are distinctly visible in the fused labels. Minor differences can be observed between the segmentation labels on atlases across the four phases, suggesting that the anatomical features and contrast properties have been maintained in the multi-phase atlases.

\begin{figure*}[t!]
    \centering
    \includegraphics[width=\textwidth]{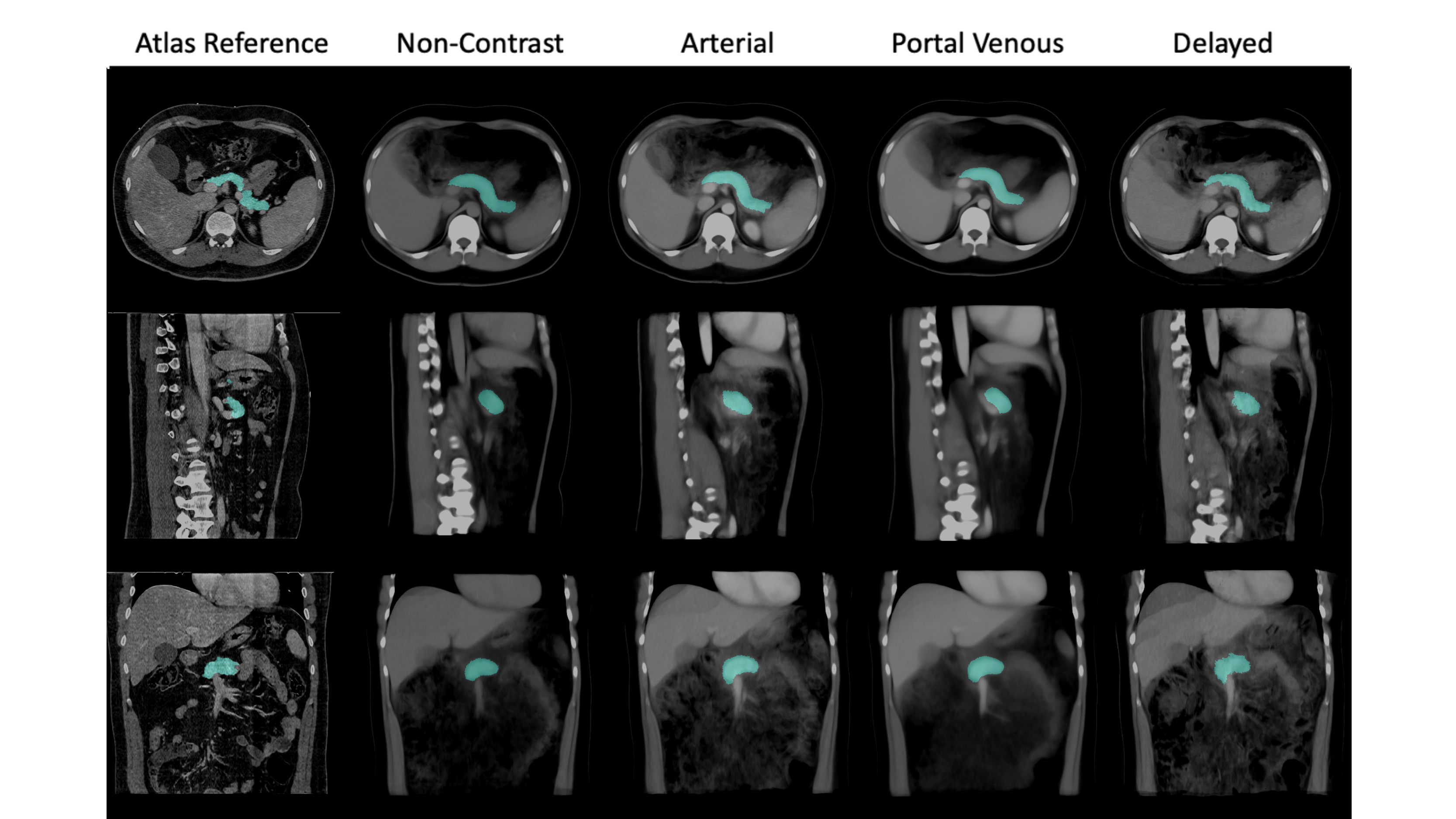}
    \caption{We investigate the registration stability across all contrast phase with average mapping. We observe that the morphology of pancreas across all phases are well preserved with clear boundaries.}
\end{figure*}
\par The fairness of the obtained atlas using the DEEDS registration method for different phases is also evaluated with inverse transform. Table 1 displays the performance of various registration tools, including ANTS, NIFTYREG, and DEEDS, to create the average atlas for the pancreas and other abdominal organs in terms of Dice score and symmetric Hausdorff distance. The performance of affine registration alone is significantly lower than that of the two-stage registration. With the two-stage registration, DEEDS achieves the highest Dice score and the lowest Hausdorff distance for the pancreas region. Another 100 subjects in the portal venous phase with 13 multi-organ ground truth segmentation labels are used to evaluate the performance of our framework, as shown in Table 2. To ensure the alignment of all organs in the atlas, we calculated the Dice score and Hausdorff distance on 100 subjects for the pancreas and the other 12 organs without qualitative assessment. DEEDS achieves the highest performance with a mean Dice score of 0.504 for all 13 organs. 

\subsection{Ablation Study}
In our ablation study, we further investigate the influence of the BPR on the performance of DEEDs registration with different range of cropping values. We evaluate five distinct cropping range to establish a similar field of interest between the moving subjects and the reference subject. The raw data without BPR offers a comprehensive view of the pelvis, abdomen, and chest regions. The cropping value of -8 to 7 encompasses parts of the pelvic area and the majority of the chest area. The cropping value of -6 to 5 excludes the pelvic region and some of the chest region. The cropping value of -4 to 5 retains solely the abdominal region, while the cropping value of -2 to 4 effectively preserves the pancreas and kidney regions but does not entirely maintain the spleen region. For the experiments without BPR, -8 to +7, and -6 to +5, we cropped only the moving subjects. For the experiments of -4 to +5 and -2 to +4, we cropped both the reference volume and the moving volume.
\par Figure 4 displays the Dice scores for 13 abdominal organs on 20 registered subjects in the portal venous phase from the BTCV public dataset. The use of body part regression significantly improves the registration performance for all cropping values. The cropping value of -6 to +5 results in the highest average Dice score on the 13 organs and the highest value on the pancreas. Consequently, we employ the -6 to +5 value to crop multi-phase subjects and compute the average atlas from them. Additionally, the value of -6 to +5 maintains a complete view of all abdominal regions, facilitating a more comprehensive analysis of abdominal organs within an anatomical context.

\begin{figure*}[t!]
    \centering
    \includegraphics[width=\textwidth]{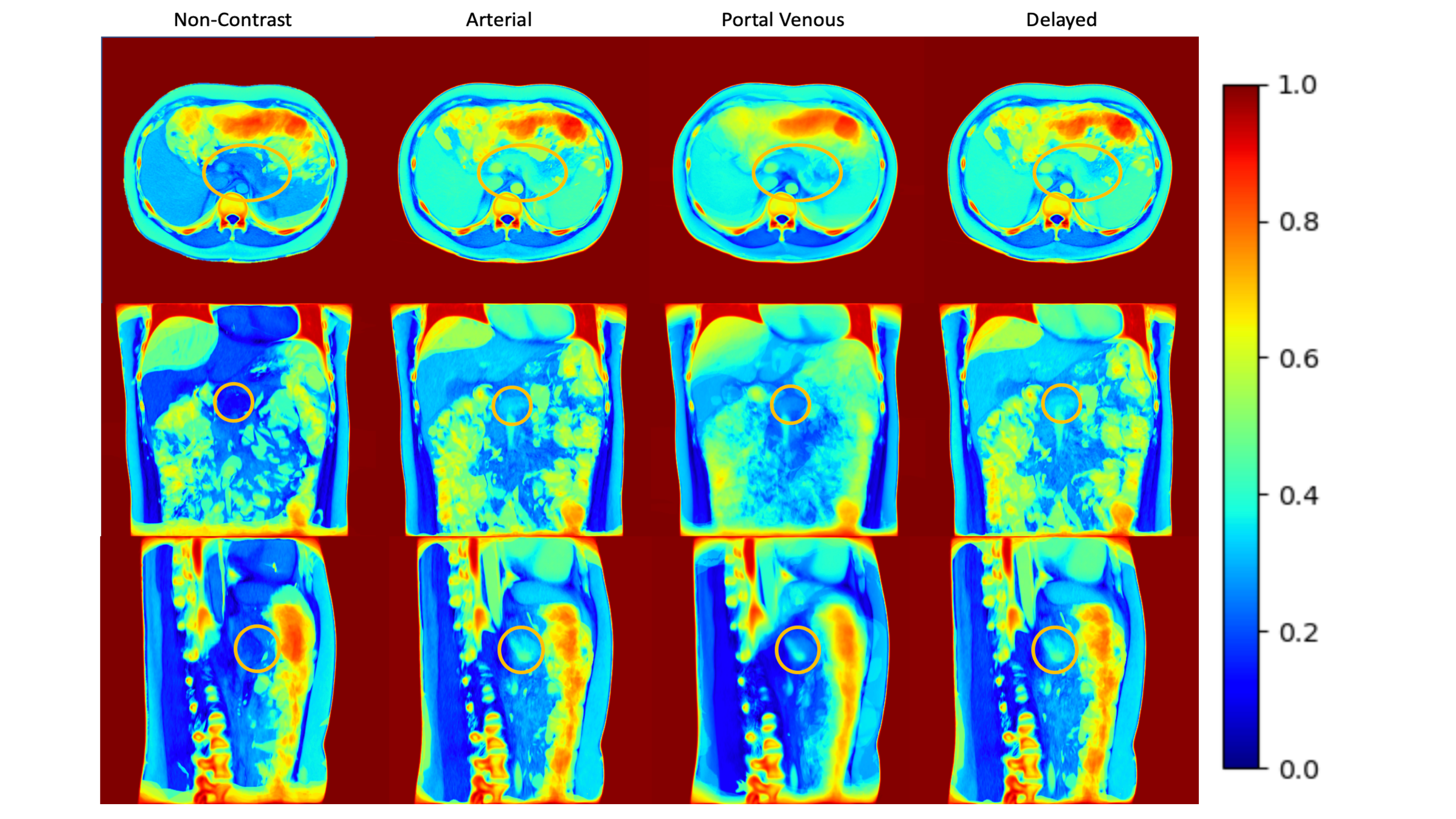}
    \caption{We further evaluate the intensity variance across the registration outputs with the average template in each contrast phase. The variance mapping in both non-contrast phase and portal venous phase demonstrates the pancreas context transferal with stability and the variance value near the pancreas region is ~0, while the range of variance value is higher in both arterial and delayed phases.}
\end{figure*}
\section{Discussion}
Population-based tissue maps play a crucial role in studying human organs by generalizing the variability between individuals. In abdominal scans, the heterogeneity of abdominal and retroperitoneal scans presents challenges in constructing an atlas capable of visualizing anatomical features and spatial relationships between organs. This study aims to propose a framework for constructing an abdominal atlas optimized for the pancreas region in multi-contrast CT.

Given the varying anatomical features across different phases of multi-contrast CT, it is essential to generalize the characteristics for each phase rather than using a single atlas template. The non-contrast phase involves acquiring images before injecting the contrast agent and serves as the baseline reference. For the pancreas, the arterial and portal venous phases are frequently used for detecting and characterizing lesions when blood vessels and abdominal organs have been further enhanced by the contrast agent.

To transfer the anatomical characteristics of each phase, we register individual subjects in different phases to the high-resolution atlas template. The robustness of registration tools is vital for the quality of the average atlas. A two-stage registration is used, where the affine registration provides prior information for the deformable registration in the second stage. As shown in Table 2, DEEDs registration achieves an average Dice score of 0.504 on 13 organs, demonstrating accurate transfer of anatomical information for all abdominal organs. ANTs is the least robust, potentially due to its surface-based registration nature. When using ANTs for label registrations, its performance improves, but 3D intensity images are still problematic due to partial matching, especially in aligning the boundaries of abdominal organs. NIFTYREG slightly outperforms ANTs, employing a block-matching approach for non-linear registration, which provides more accurate results in the presence of large deformations but requires longer computation time. DEEDs, a voxel-based method, surpasses the other two methods in performance, possibly because it relies on discrete optimization, allowing greater control over the displacement space \cite{heinrich2013mrf}. One advantage of DEEDs is its use of dense stochastic sampling approaches, sampling random voxels from non-overlapping cubes, and then calculating the displacement on cubes. It ensures accurate registration of small anatomical features undergoing large motion. Additionally, discrete optimization reduces computational complexity, making it more efficient than continuous optimization.

Nonetheless, registration accuracy is still not ideal and could influence the quality of the average atlas. Challenges for these registration tools include 1) the dissimilar field of view between moving subjects and the target template, and 2) the variation of secondary structures (e.g., muscles, bones) across the population. Many registration failures occur due to mismatched fields of view between subjects, necessitating better pre-processing steps. In the ablation study, we demonstrated the effectiveness of body part regression in matching the field of view. Secondary structures could also distract from the registration of target organs. The variation of these parts could cause undesirable deformation, especially when they occupy a large space in abdominal scans\cite{xu2016evaluation}. Similarly, small or medium-sized organs could be affected by surrounding organs. For instance, pancreas deformation might be negatively affected by other surrounding target organs with larger sizes, such as the spleen.

In addition to traditional optimization-based registration, learning-based registration methods have gained popularity due to their fast speed. However, concerns remain regarding the use of learning-based methods for constructing average maps in our study. Optimization-based models generally exhibit better expressive power because model parameters are optimized for a specific pair of images, enabling sharper deformation to preserve the details of anatomical features of abdominal organs. On the other hand, learning-based methods optimize the model for the entire dataset, which tends to produce over-smoothed transformations, potentially losing individual characteristics during registration and making them less suitable for high-resolution images \cite{nazib2018comparative}. Furthermore, learning-based methods typically exhibit less generalizability and are more specific to the dataset.

In addition to constructing average atlases, fused labels on the atlas provide clearer anatomical characteristics and contextualization of the pancreas. As demonstrated in Figure 6, the shape and positions of average segmentation labels closely resemble those in the atlas template. However, limitations still exist. Since the segmentation represents the average of subjects, it is not specific to individual subjects; thus, detailed features and irregular boundary may not be accurately represented on the fused labels, providing only an estimated shape of the pancreas. Another concern regarding segmentation labels on atlases is the accuracy of the segmentation performance of the model. We employed UNesT, which was trained on the BTCV dataset in the portal venous phase. However, due to minor differences in anatomical features between subjects in each phase, segmentation in other phases may not be as accurate as in the portal venous phase.

\section{Conclusion}
This study introduces a high-resolution pancreas atlas framework to generalize the healthy biomarker across popluation with multi-contrast abdominal CT. By utilizing body part regression to match the field of view between the target template and moving subjects, and employing the DEEDs registration method to transfer subjects into the target space, our atlas effectively captures population-wide features of the pancreas organ and contextualizes the anatomical characteristics of the pancreas within the entire abdominal scans. Future work involving the use of pancreas atlases could further explore areas such as enhancing the segmentation accuracy of the pancreas and improving the localization of the pancreas in the context of pathological changes. \\

%
%
%
%
\noindent{\bf Acknowledgements} This research is supported by NIH Common Fund and National Institute of Diabetes, Digestive and Kidney Diseases U54DK120058 (Spraggins), NSF CAREER 1452485, NIH 2R01EB006136, NIH 1R01EB017230 (Landman), and NIH R01NS09529. This study was in part using the resources of the Advanced Computing Center for Research and Education (ACCRE) at Vanderbilt University, Nashville, TN. The identified datasets used for the analysis described were obtained from the Research Derivative (RD), database of clinical and related data. The imaging dataset(s) used for the analysis described were obtained from ImageVU, a research repository of medical imaging data and image-related metadata. ImageVU and RD are supported by the VICTR CTSA award (ULTR000445 from NCATS/NIH) and Vanderbilt University Medical Center institutional funding. ImageVU pilot work was also funded by PCORI (contract CDRN-1306-04869).

\bibliographystyle{splncs04}
\bibliography{MyLibrary}

\end{document}